\documentclass[12pt]{article}%
\usepackage{amsmath,latexsym}
\usepackage{graphicx}
\usepackage{amsmath}
\usepackage{amsfonts}
\usepackage{amssymb}%
\providecommand{\U}[1]{\protect\rule{.1in}{.1in}}
\begin{document}

\title{Stability of Non-asymptotically flat thin-shell wormholes   in generalized dilaton-axion gravity}
\author{Ayan Banerjee\thanks{%
ayan\_7575@yahoo.co.in}, Farook Rahaman\thanks{%
 rahaman@iucaa.ernet.in},
Surajit Chattopadhyay\thanks{%
surajit\_2008@yahoo.co.in},
 Sumita Banerjee\thanks{%
banerjee.sumita.jumath@gmail.com}  \and$^{\ast}$ {\small
Department of Mathematics, Jadavpur University, Kolkata - 700032,
India }  \and $^{ \dag}$ {\small Department of Mathematics,
Jadavpur University, Kolkata - 700032, India }\and $^{\ddag }$
 {\small Pailan College of Management and Technology,
Bengal Pailan Park, Kolkata-700104, India}  \and  $^{\S }$
{\small
 Adamas Institute of Technology, Barasat,
North 24 Parganas - 700126, India} } \maketitle

\date{}

\begin{abstract}
 We construct a new type of thin-shell wormhole for
non-asymptotically flat charged black holes in generalized
dilaton-axion gravity inspired by low-energy string theory using
cut-and-paste technique. We have shown that this thin shell
wormhole is stable. The most striking feature of our model is that
the total amount of exotic matter needed to support the  wormhole
can be reduced as desired with the suitable choice of the value of
a parameter. Various other aspects of thin-shell wormhole are also
analyzed.
\end{abstract}

\maketitle

\section{Introduction}

   ~~ The study of traversable wormholes and thin-shell wormholes are very
   interesting subject  in recent years
as it is known that wormholes represent shortcut in space-time,
though the observational evidence of wormholes has not been
possible still now.
 At first, Morris and Thorne \cite{MT1988} elaborated the structure of traversable wormhole with two mouths and a throat.
 The traversable wormhole was obtained as a solution of Einstein's equations having two asymptotically flat regions.
  These flat regions are connected by a minimal surface area, known as the throat satisfying the flare-out condition
  \cite{HV1997}. This type of wormhole would allow the travel between two parts of the same universe or between two different universes.
  Morris and Thorne \cite{MT1988} had the idea to make the conversion of a wormhole traversing space into
   a wormhole traversing time.
 In the Morris and Thorne model of wormhole  the violation of positive energy condition is unavoidable. The throat of the  wormhole
  is held open with exotic matter.~  Since it is very difficult to deal with the exotic matter
(violation of positive energy condition) Visser \cite{v1989},
 proposed a way to minimize the usage of exotic matter by
applying cut-and-paste technique on a black hole to construct a
new class  of spherically symmetric wormholes, known as thin-shell
wormhole in which the  exotic matter is concentrated within the
joining shell. This shell can act as the    wormhole throat. Using
the Darmois-Israel formalism\cite {Israel1966}, the surface
stress-energy tensor components within the shell i.e.  at the
throat could be determined.   Visser's approach is very simple for
theoretical construction   of wormhole and  perhaps also practical
because it minimizes the amount of exotic matter required and
therefore this approach was adopted by various authors to
construct thin shell wormholes
\cite{Poisson1995,Lobo2003,Lobo2004,Eiroa2004a,Eiroa2004b,Eiroa2005,
Thibeault2005,Lobo2005,Rahaman2006,
Eiroa2007,Rahaman2007a,Rahaman2007b,Rahaman2007c,
Lemos2007,Richarte2008,Rahaman2008a,
Rahaman2008b,Eiroa2008a,Eiroa2008b,Rahaman2010a, Rahaman2010b,
Dias2010,Peter2010,fa12a,e12,p12,fa11}.

Recently, Mazharimousavi et al \cite{MH} have studied
d-dimensional non-asymptotically flat thin-shell wormholes in
Einstein-Yang-Mills-Dilaton gravity where as Rahaman et al
\cite{fa12b} have obtained a new class of stable (2+1) dimensional
non-asymptotically flat thin-shell wormhole.    In this
investigation, we are searching stable non-asymptotically flat
thin-shell wormholes in generalized dilaton-axion gravity.

In 2005,   Sur,   Das and   SenGupta\cite{Sur2005} discovered new
black hole solutions considering gravity coupled with dilaton
($\varphi$), Kalb-Ramond axion ($\zeta$) scalar fields and a
Maxwell field with arbitrarily coupled to the scalars. They
considered the generalized action in four dimensions as:
\begin{equation}
\mathcal{I}=\frac{1}{2\kappa}\int
d^4x\sqrt{-g}[R-\frac{1}{2}\partial_{\mu}\varphi\partial^{\mu}\varphi
-\frac{\omega(\varphi)}{2}\partial_{\mu}\zeta\partial^{\mu}\zeta-\alpha(\varphi,\zeta)F^2-\beta(\varphi,\zeta)F_{\mu\nu}\ast
F^{\mu\nu}].
\end{equation}
where $\kappa = 8\pi G$, R is the curvature scalar and $F_{\mu\nu}$ is the Maxwell
field tensor where two massless scalar or
pseudo scalar fields are represented by $\varphi$ and
$\zeta$  which are coupled to the Maxwell
field with the functional dependence by the functions $\alpha(\varphi,\zeta)$and
$\beta(\varphi,\zeta)$.

  Using the above action,  Sur,Das and SenGupta\cite{Sur2005} have found
  two types
of black-hole solutions, classified as asymptotically flat and
asymptotically non-flat. For our thin-shell wormhole solution, we
have considered  non asymptotically flat metric given by
\begin{equation}
ds ^2 = - f(r) dt^2 + f(r)^{-1}dr^2 + h(r) d\Omega^2.\label{eq22}
\end{equation}
where
\begin{equation}
 f(r) =\frac{(r-r_+)(r-r_-)}{r^2\left(\frac{2r_0}{r}\right)^{2n}},
\end{equation}
and
\begin{equation}
h(r)=r^2\left(\frac{2r_0}{r}\right)^{2n}.
\end{equation}

The parameter  n is a dimensionless constant lies in the range $ 0
<n<1$.
 The various parameters for non asymptotically flat metric are given by
\begin{equation}
r_{\pm}=\frac{1}{(1-n)}\left[m\pm\sqrt{{m^2}-{(1-n)\frac{K_2}{4}}}\right],
\end{equation}

\begin{equation}
r_0=\frac{1}{16m_0}\left(\frac{K_1}{n}-\frac{K_2}{1-n}\right),
\end{equation}
\begin{equation}
m_0=m-\left(2n-1\right)r_0,
\end{equation}
\begin{equation}
K_1=16nr_0^2,
\end{equation}
\begin{equation}
K_2=4(1-n)r_{+}r_{-},
\end{equation}
\begin{equation}
m=\frac{1}{16r_0}\left(\frac{K_1}{n}-\frac{K_2}{1-n}\right)+\left(2n-1\right)r_0.
\end{equation}
where m is the mass of the black hole and the inner, outer event
horizons and curvature singularity represented by the parameters
$r_+$ ,$r_-$ and $r = r_0$ respectively. The parameters obey the
restriction $r_0 < r_- < r_+$.
\\
Employing the above black hole solution, we have constructed a new
thin-shell wormhole. We have shown that this thin shell wormhole
is stable. The most striking feature of our model is that the
total amount of exotic matter needed to support the  wormhole can
be reduced as desired by choosing the  parameter n very close to
unity. Various other aspects of thin-shell wormhole have  also
been  analyzed.

  ~~ We have organized the work as follows: In
section 2,
 we have provided mathematical formulation for constructing  a thin-shell wormhole from
charged black hole in generalized dilaton-axion gravity. In
section 3, we have discussed the effect of gravitational field on
wormhole i.e. attractive or repulsive nature of the wormhole. In
section 4, we have discussed about the equation of state relating
pressure and density at the wormhole throat and in section 5,  we
have calculated  total amount of exotic matter. Linearized
stability analysis has been discussed in section 6. Finally, we
have made a conclusion about the work.

\section{ Construction of thin-shell wormhole}
 For the construction of thin-shell wormhole, at first, we remove
two identical copies from four-dimension spacetime of the black
hole given in equation (2) region with radius $r \leq a$. Here we
have assumed  $a>r_{+}$ to avoid all types of singularity. Thus we
have two copies of regions:
\begin{equation}
\mathcal{M^{\pm}}\equiv\lbrace r=a:a>r_{+} \rbrace.
\end{equation}
Now, we  stick them together at the hypersurface,
$\Sigma=\Sigma^{\pm}$, to get a geodesically complete manifold $ M
= M^+ \bigcup M^- $. Here,  two regions are connected by minimal
surface area, called throat with radius a.

  The induced 3-metric
on the hypersurface is given by
\begin{equation}
ds ^2 = - d\tau^2 + h[r(\tau)]d\Omega^2.\label{eq22}
\end{equation}
where $\tau$,represents the proper time on the junction surface.
To obtain the surface stress-energy tensor
$S^{i}_{j}=diag\left(-\sigma,P,P\right)$, we use the Lancozos
equations \cite {Israel1966}, which reduce to

 \begin{equation}
\sigma=-\frac{1}{4\pi}\frac{h^{\prime}(a)}{h(a)}\sqrt{f(a)+\dot{a}^2},
\end{equation}
and
\begin{equation}
P_{\theta}=P_{\phi}=P=\frac{1}{8\pi}\frac{h^{\prime}(a)}{h(a)}\sqrt{f(a)+\dot{a}^2}+\frac{1}{8\pi}\frac{f^{\prime}(a)+2\ddot{a}}{\sqrt{f(a)+\dot{a}^2}}.
\end{equation}
where $\sigma$ and $P$ represents the surface energy density and
surface pressures respectively. To understand the dynamics of the
wormhole ,we consider that the throat as function of proper time
i.e.$a=a(\tau)$ and  over dot and prime denote the derivatives
with respect to $\tau$ and a. For a static configuration of radius
a (assuming $\dot a=0$ and $ \ddot a=0$), we obtain the values of
the surface energy density and the surface pressures as

\begin{equation}
\sigma=-\frac{(1-n)(a-r_{+})(a-r_{-})}{a^{2-n}\mathcal{K}},
\end{equation}
\begin{equation}
P=\frac{2a-r_{+}-r_{-}}{4\mathcal{K} a^{1-n}},
\end{equation}
where
\begin{equation}
\mathcal{K}=2\pi\left(2r_0\right)^{n}\sqrt{(a-r_{+})(a-r_{-})}.
\end{equation}
From Eqs.(15) and (16), we see that the energy density $\sigma$ is
negative, however the pressure p is positive depending on the
position of the throat and on the parameters $r_0$, $r_{-}$ and
$r_{+}$ defining on wormhole. Here, matter distribution within the
shell  of the wormhole violates the weak energy condition.

  ~~ We plots $\sigma$ and P versus a in Figs.(1-2)
for such wormholes whose radii fall within the range of 6 to 16
km, keeping the restriction on the parameters $r_{+} > r_{-} >
r_{0}$. The sensitivity are given with respect to n, for both the
figures $\sigma$ and P as described in the caption of the figures.

Note that the weak energy condition requires that  $\sigma >0$ and
$\sigma +P >0$. These state that the energy density is positive
and the pressure is not too large compared to the energy density.
The null energy condition $\sigma +P >0$ is a special case of the
latter and implies that energy density can be negative if there is
a compensating positive pressure. The figure 3 indicates in the
present case that the null energy condition is satisfied.

\begin{figure}[ptb]
\begin{center}
\vspace{0.2cm}
\includegraphics[width=0.6\textwidth]{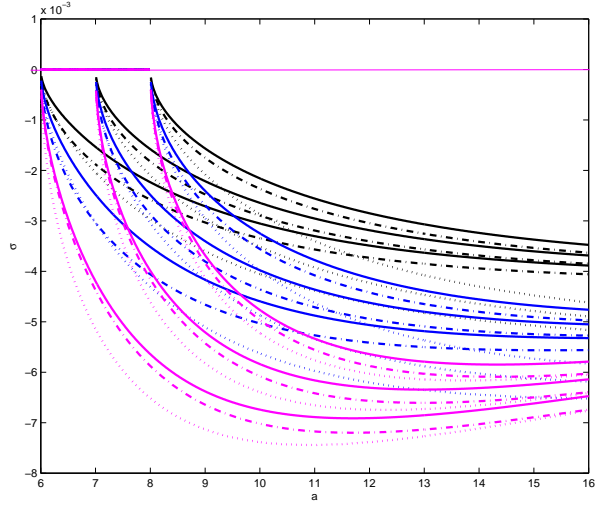}
\end{center}
\caption{For every branch of curves, colors black, blue and pink
represent for n = 0.7, 0.5 $\&$ 0.1 and $r_{+}$ = 8, 7 $\&$  6
respectively. For every combination of n and $r_{+}$, we draw
three different sets when ($r_{-}$ = 5, $r_{0}$ = 3), ($r_{-}$ =
5, $r_{0}$ = 2)and ($r_{-}$ = 4, $r_{0}$ = 3) which are shown by
solid, dot-dash and dotted curves  respectively.} \label{fig5}
\end{figure}
\begin{figure}[ptb]
\begin{center}
\vspace{0.2cm}
\includegraphics[width=0.6\textwidth]{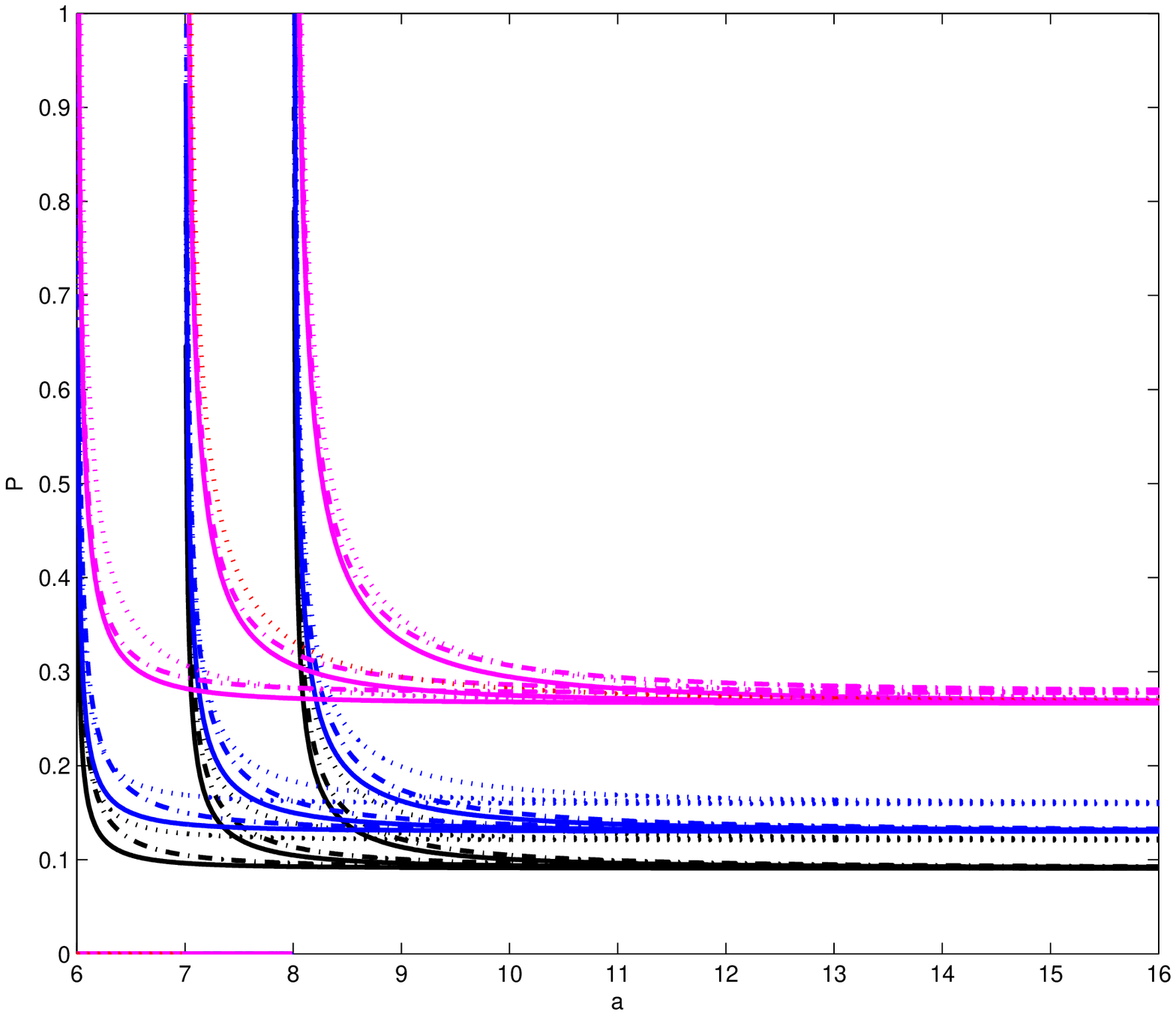}
\end{center}
\caption{For the branch of curves, the description of the figure is same as in Fig1.   } \label{fig5}
\end{figure}
\begin{figure}[ptb]
\begin{center}
\vspace{0.2cm}
\includegraphics[width=0.6\textwidth]{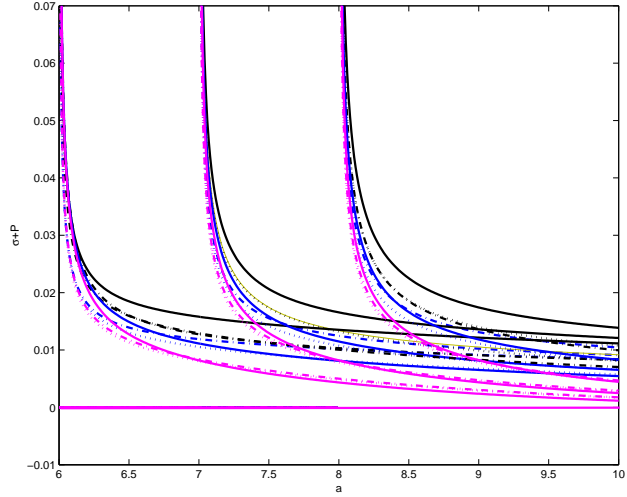}
\end{center}
\caption{For the branch of curves, the description of the figure
is same as in Fig1.   } \label{fig5}
\end{figure}

\section{The gravitational field}

In this section we have  analyzed the  nature of the gravitational
field of the wormhole constructed from charged black holes in
generalized dilaton-axion gravity and  calculate the observer's
four-acceleration $A^\mu = u^\mu_{\,\,;\nu} u^\nu$, where $u^{\nu}
=  d x^{\nu}/d {\tau} =(1/\sqrt{f(r)}, 0,0,0)$.  Only non-zero
component for the line element in Eq.(1), is given by

\begin{equation}
A^r = \Gamma^r_{tt} \left(\frac{dt}{d\tau}\right)^2 \\=
\frac{r^{2n-3}}{(2r_0)^{2n}}\left[nr^2-\left(n-\frac{1}{2}\right)(r_{+}
+r_{-})r +(n-1)r_{+}r_{-}\right].
\end{equation}
 The equation of motion of a  test particle  is given by
\begin{equation}
\frac{d^2r}{d\tau^2}= -\Gamma^r_{tt}\left(\frac{dt}{d\tau}
\right)^2 =-A^r.
\end{equation}
when it is radially moving and  initially at rest.

For $A^r =0$,  we get the geodesic equation. From the figure 4, we
note
 that the wormhole is attractive as $A^r>0$ out side the event
 horizon.
\begin{figure}[ptb]
\begin{center}
\vspace{0.2cm}
\includegraphics[width=0.5\textwidth]{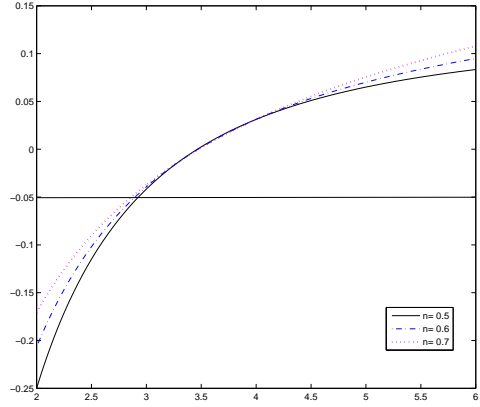}
\end{center}
\caption{ We draw the figure for the acceleration $A^r$ (vertical
axis)  with respect to $a$ (horizontal axis) for different values
of n assuming $r_{+}=4$,$r_{-}=3$ and $r_{0}=2$. Figure indicates
attractive nature of the wormhole as $A^r>0$ for $a>r_+$.  }
\label{fig5}
\end{figure}

\section{An equation of state}

 To know the exact nature of the matter distribution in the shell,  we calculate
 equation of state (EoS) which is given by
\begin{equation}
\frac{P}{\sigma}=w=-\frac{a\left(2a-r_{+}-r_{-}\right)}{4(1-n)(a-r_{+})(a-r_{-})}.
\end{equation}
The plot for w (fig 5)  indicates that  matter distribution in the
shell is phantom energy type.
  For  $a\rightarrow \infty$ i.e. if
the wormhole throat is very very large,  then $w \rightarrow
-\frac{1}{2(1-n)}$. In that case, EoS solely depends on the
parameter n only. One can note that for the following range of n,
$\frac{1}{2}<n<\frac{3}{2}$ , w lies on $-1<w<-\frac{1}{3}$. This
indicates the mater distribution is quentessence like. However,
for $n>\frac{3}{2}$, $w<-1$ i.e. the matter distribution is
phantom energy type as $w<-1$.
\begin{figure}[ptb]
\begin{center}
\vspace{0.2cm}
\includegraphics[width=0.5\textwidth]{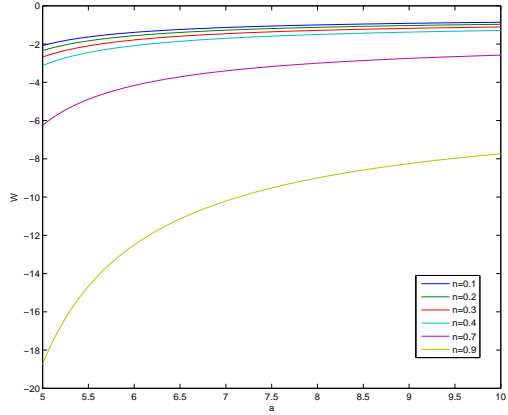}
\end{center}
\caption{ We draw the figure for w with different values of n
assuming $r_{+}=4$, $r_{-}=3$. Figure indicates the matter
distribution is phantom energy type.} \label{fig5}
\end{figure}

\section{The total amount of exotic matter}

 We observe that matter distribution at throat
is phantom energy type i.e. matter at throat is exotic. Here, we
evaluate the total
 amount of exotic matter which can
be quantified by the integral
\cite{Eiroa2005,Thibeault2005,Lobo2005,Rahaman2006,
Eiroa2007,Rahaman2007a,Rahaman2007b}

\begin{equation}
   \Omega_{\sigma}=\int [\rho+p ]\sqrt{-g}d^3x.
\end{equation}
where $g$ represents the determinant of the metric tensor.

Now, by using the radial coordinate $R=r-a$, we have
\begin{equation}
 \Omega_{\sigma}=\int^{2\pi}_0 \int^{\pi}_0 \int^{\infty}_{-\infty}
     [\rho+p]\sqrt{-g}\,dR d\theta d\phi.
\end{equation}
For the infinitely thin shell it does not exert any radial
pressure and using  $\rho=\delta(R)\sigma(a)$
we have,
\begin{equation}\label{E:amount}
 \Omega_{\sigma}=\int^{2\pi}_0 \int^{\pi}_0 \left.[\rho\sqrt{-g}]
   \right|_{r=a} d\theta d\phi=4\pi h(a)\sigma(a)
=-\frac{\left(1-n\right)\mathcal{K}}{a^n}.
\end{equation}
where $\mathcal{K}$ is given in Eq.(17).

It is to be noted that the total  amount of exotic matter needed
can be reduced as small as desired by choosing the parameter n
very close to unity (see figure 6). Also, the exotic  matter can
be minimized  by taking the value of $a$ very close to the
location of the outer event horizon (see figure 7).

\begin{figure}[ptb]
\begin{center}
\vspace{0.2cm}
\includegraphics[width=0.5\textwidth]{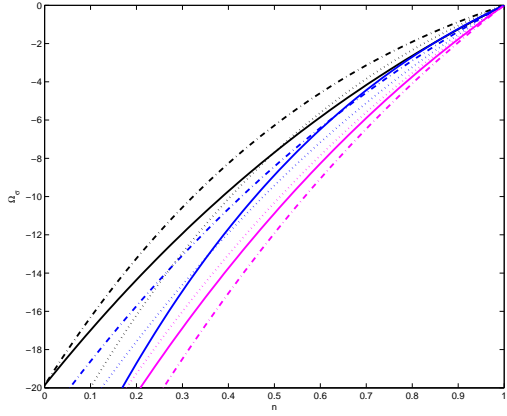}
\end{center}
\caption{ The plot for $\Omega_{\sigma}$ versus n, for fixed value
of a = 10. For the branch of curves, the description of the figure
is same as in Fig1.    } \label{fig5}
\end{figure}

\begin{figure}[ptb]
\begin{center}
\vspace{0.2cm}
\includegraphics[width=0.5\textwidth]{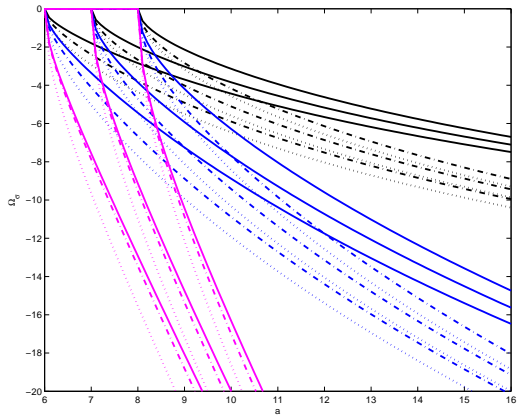}
\end{center}
\caption{ The plot for $\Omega_{\sigma}$ versus a. For the branch
of curves, the description of the figure is same as in Fig1.   }
\label{fig5}
\end{figure}

\section{Linearizing method}

Now, we are trying to find the local stability of the
configuration under small perturbation around the static solution
at $a = a_0$. So, rearranging the Eq.(13) we can write
\begin{equation}
\dot a^2+ V(a) = 0,
\end{equation}
which represents the equation of motion
of the shell,
where the potential $V(a)$ is define as
\begin{equation}
V(a) = f(a)-16\pi^{2}\biggl[\sigma(a)\frac{h(a)}{h^{\prime}(a)}\biggr]^{2}.
\end{equation}
Using the Taylor series expansion for $V(a)$ around the static
solution $a_0$, we get
\begin{eqnarray}
V(a) &=&  V(a_0) + V^\prime(a_0) ( a - a_0) +
\frac{1}{2} V^{\prime\prime}(a_0) ( a - a_0)^2  \nonumber \\
&\;& + O\left[( a - a_0)^3\right],
\end{eqnarray}
where the prime denotes the derivative with respect to a.

  ~~For our stability analysis, we start with the energy conservation
equation. Now using Eqs.(15) and (16), preserving the symmetry of
linearizing radial perturbations,  the energy conservation
equation defined as
\begin{equation}
\frac{d}{d\tau}(\sigma \mathcal{A})+P\frac{d \mathcal{A}}{d \tau} = \lbrace [h^{\prime}(a)]^{2}
 -2h(a)h^{\prime\prime}(a)\rbrace  \frac{\dot{a}\sqrt{f(a)+\dot{a}^2}}{2h(a)},
\end{equation}
where $\mathcal{A}=4\pi h(a)$ denotes the area of the wormhole
throat.

 From Eq.(25), one can get

\begin{equation}
\frac{d}{da}[\sigma h(a)]+P\frac{d}{d a}[h(a)] = -\lbrace [h^{\prime}(a)]^{2}
 -2h(a)h^{\prime\prime}(a)\rbrace  \frac{\sigma}{2h^{\prime}(a)},
\end{equation}
which finally gives
\begin{equation}
h(a)\sigma^{\prime}+h^{\prime}(a)(\sigma+P)+\lbrace [h^{\prime}(a)]^{2}
 -2h(a)h^{\prime\prime}(a)\rbrace  \frac{\sigma}{2h^{\prime}(a)}=0,
\end{equation}

The second order derivative for the potential  can be defined as

\begin{eqnarray}
V^{\prime\prime}(a) &=& f^{\prime\prime}(a)+16\pi^{2}\times \Bigg\lbrace \biggl[\frac{h(a)}{h^{\prime}(a)}\sigma^{\prime}(a)+\left(1-\frac{h(a)h^{\prime\prime}(a)}{[h^{\prime}(a)]^2}\right)\sigma(a)\biggr]\nonumber \\
&\;& \times[\sigma(a)+2P(a)]+ \frac{h(a)}{h^{\prime}(a)}\sigma(a)[\sigma^{\prime}(a)+2p^{\prime}(a)]\Bigg\rbrace  ,
\end{eqnarray}
~~ Now, using the parameter $\beta^{2}=\frac{d P}{d \sigma}$,
which is normally interpreted as the speed of sound,  the above
expression can be written as
\begin{eqnarray}
V^{\prime\prime}(a) &=& f^{\prime\prime}(a)-8\pi^{2}\times \Bigg\lbrace [\sigma(a)+2P(a)]^{2}\nonumber \\
&\;& +2\sigma(a)\bigg[\left(\frac{3}{2}-\frac{h(a)h^{\prime\prime}(a)}{[h^{\prime}(a)]^2}\right)\sigma(a)+P(a)\biggr](1+2\beta^{2})\Bigg\rbrace ,
\end{eqnarray}
 The solution gives a stable configuration if $V(a)$ process a local minima at$a_0$, in other words,
  $V''(a_0)>0$. Now using the condition $V(a_0)=0$ and $V'(a_0)=0$, we  solve
for $\beta^{2}$ as
\begin{equation}
\beta^{2}=-\frac{1}{2}+\frac{\frac{f''}{8\pi^{2}}-(\sigma+2P)^2}{4\sigma\bigg[\left(\frac{3}{2}-\frac{hh^{\prime\prime}}{[h^{\prime}]^2}\right)\sigma+P\biggr]}.
\end{equation}
Now,  we are trying find the stable region with the  help of
graphical representation due to complexity of the expression of
$V''(a_0)>0$. In Figs. 8-11, we
 find the possible range of $a_0$, where $V(a_0)=0$
possess a local minima.  We plot the figures 8 and 9 for the
parameters $r_{0}=2$, $r_{-}=3$ and $r_{+}=4$ and different values
of n=0.5 and 0.7, which show that graphs don't possess any stable
configuration as ${\beta}^2$ lies outside (0,1).  However, if we
choose the the parameters $r_{0}=2$, $r_{-}=1$, $r_{+}=6$ and
n=0.5, respectively, then the stable region is more closer to the
normal range of ${\beta}^2$. This indicates  that the increase of
the difference between inner and outer event horizons, stable
region may fall  to the normal range of $\beta^{2}$. Note that
this criteria holds when we are dealing with real matter. However,
according Poisson and Visser \cite{Poisson1995} the interpretation
of ${\beta}^2$ should be relaxed  when dealing with exotic matter.
As a result, the values of ${\beta}^2$ may fall out side (0,1).
Since the stability criteria is $V''(a_0)>0$ and our figures 8-11
indicate the stable regions where $V''(a_0)>0$, therefore, our
thin shell wormholes are stable.

\begin{figure}[ptb]
\begin{center}
\vspace{0.2cm}
\includegraphics[width=0.5\textwidth]{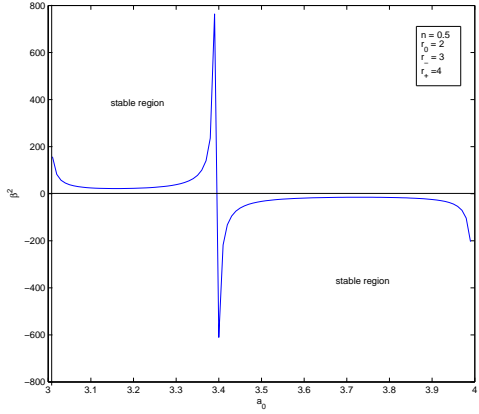}
\end{center}
\caption{The stability region is above the curve on the left
and below the curve on the right for n=0.5.   } \label{fig5}
\end{figure}

\begin{figure}[ptb]
\begin{center}
\vspace{0.2cm}
\includegraphics[width=0.5\textwidth]{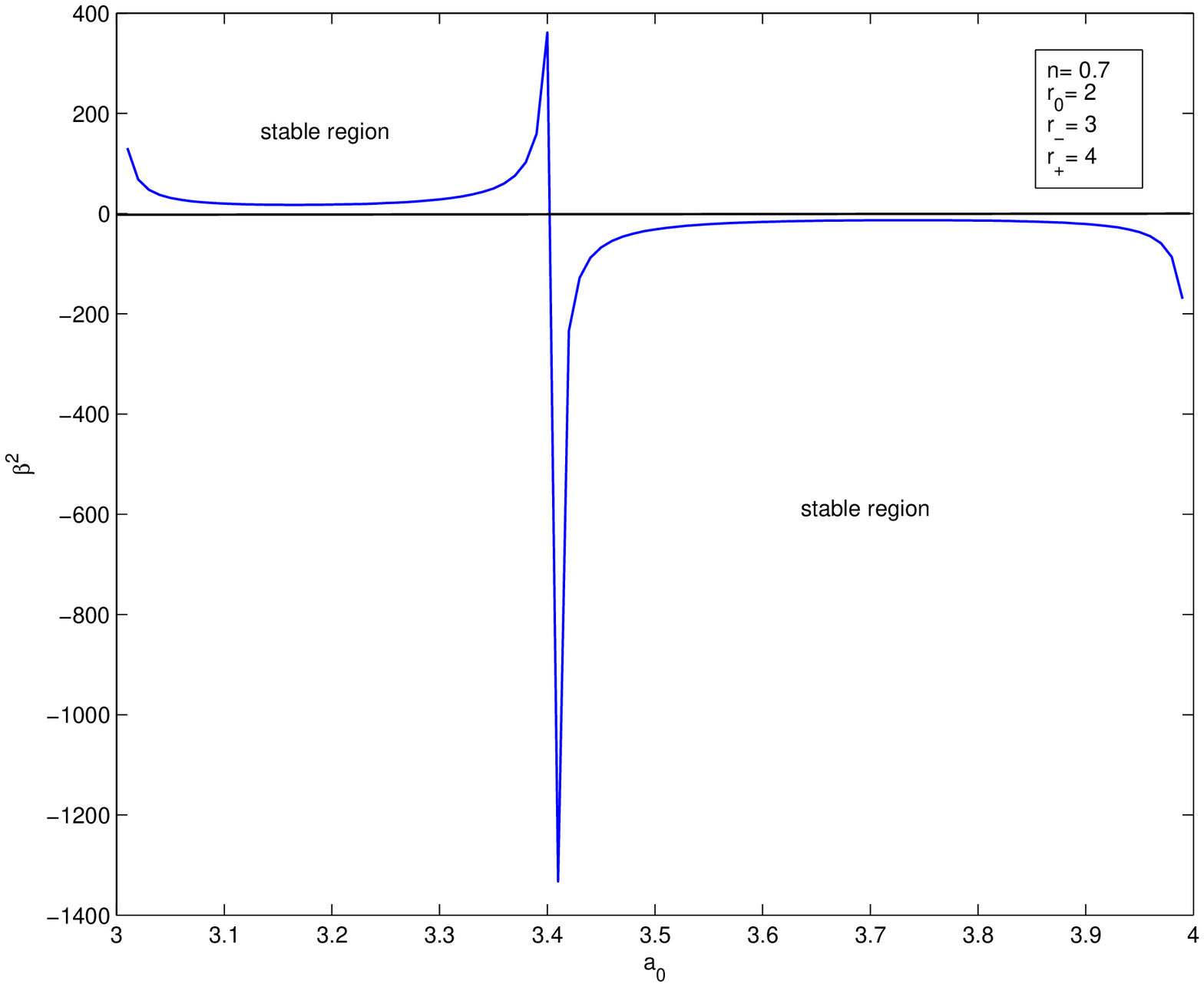}
\end{center}
\caption{The stability region is above the curve on the left
and below the curve on the right for n=0.7.   } \label{fig5}
\end{figure}
\begin{figure}[ptb]
\begin{center}
\vspace{0.2cm}
\includegraphics[width=0.6\textwidth]{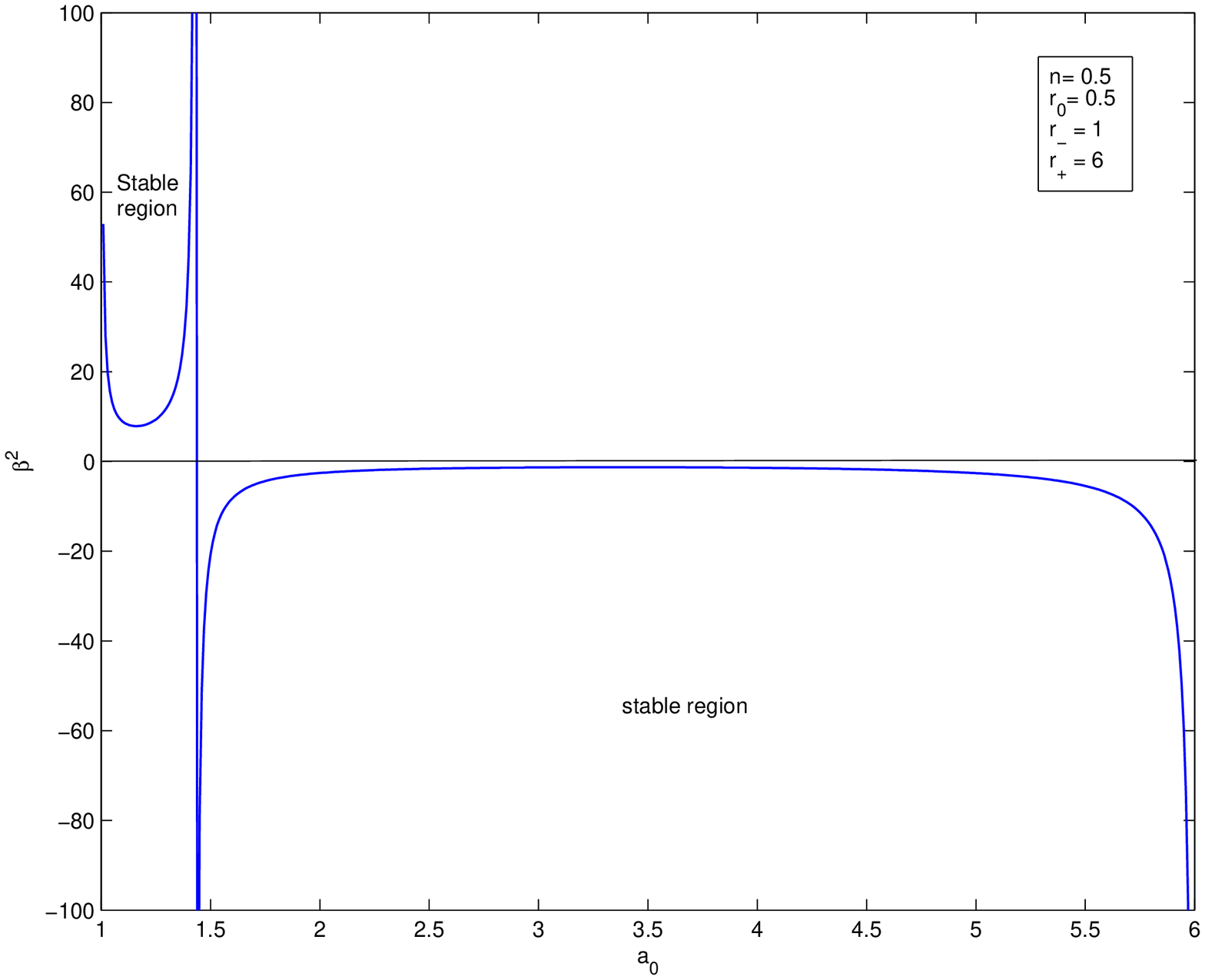}
\end{center}
\caption{ The stability region is above the curve on the left and
below the curve on the right when n=0.5 and $r_{0}=0.5$, $r_{-}=1$
and $r_{+}=6$, i.e. plots for large difference between inner and
outer event horizon.  } \label{fig5}
\end{figure}
\begin{figure}[ptb]
\begin{center}
\vspace{0.2cm}
\includegraphics[width=0.6\textwidth]{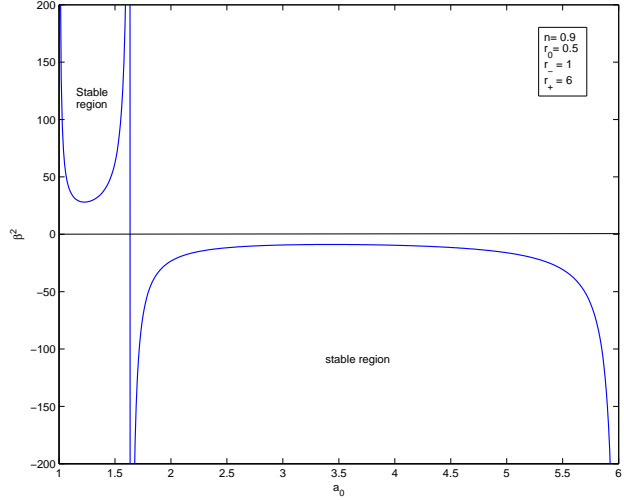}
\end{center}
\caption{ The stability region is above the curve on the left and
below the curve on the right when n=0.9 and $r_{0}=0.5$, $r_{-}=1$
and $r_{+}=6$, i.e. plots for large difference between inner and
outer event horizon.  } \label{fig5}
\end{figure}

\pagebreak

\section{Conclusions}\noindent
In this work,  we have provided a new type of thin-shell wormhole
applying the cut-and-paste technique on charged black hole in
generalized dilaton-axion gravity. The metric which we have used
is not asymptotically flat, therefore, the the wormholes are not
asymptotically flat. The matter confined within the shell of the
wormhole  violates the weak energy condition. The matter
distribution in the shell is of phantom energy type. However, when
the wormhole throat is very very large, then  for the following
range of n,  $\frac{1}{4}<n<\frac{3}{4}$,   mater distribution is
quentessence like. It is interesting to  note that the total
amount of exotic matter needed can be reduced as small as desired
by choosing the parameter n very close to unity. Finally, to show
the stability, we have performed linearized stability analysis
around the static solution. Since the matter distribution within
the shell is exotic type, therefore, according to Poisson and
Visser \cite{Poisson1995}, we can relax the range of ${\beta}^2$.
Here the stability regions are shown graphically.

\section*{Acknowledgments}   FR  is thankful to IUCAA, Pune, India for providing research facility. FR
is also grateful UGC for providing financial support.

\end{document}